\documentstyle[psfig]{mn}
\title{Peculiar Velocities and the Mean Density Parameter}
\author [Nelson Padilla and Diego G. Lambas]
	{Nelson Padilla$^1$ and Diego G. Lambas$^1$.\\
	$^1$ Grupo de Investigaciones en Astronomia Teorica y Experimental,
	IATE,\\ 
	Observatorio Astronomico, Laprida 854, 5000 Cordoba Argentina}

\date{\today}

\begin{document}
\maketitle

\begin{abstract}
We study the peculiar velocity field inferred from the
Mark III spirals using a new method of analysis.  We estimate 
optimal values of Tully-Fisher scatter and zero-point offset, 
and we derive 
the 3-dimensional rms peculiar velocity ($\sigma_v$) 
of the galaxies in the samples analysed. 
We check our statistical analysis using mock catalogs derived from 
numerical simulations of CDM models considering 
measurement uncertainties and sampling variations.  
Our best determination for the observations is
$\sigma_v= (660\pm50) km/s$.
We use the linear theory relation between $\sigma_v$ , the density
parameter $\Omega$, and the galaxy correlation function $\xi(r)$ 
to infer the quantity $\beta =\Omega^{0.6}/b = 0.60^{+0.13}_{-0.11}$ 
where $b$ is the linear bias parameter of optical galaxies and
the uncertainties correspond to bootstrap resampling and
an estimated cosmic variance added in quadrature.  Our
findings are consistent with the results
of cluster abundances and redshift space distortion of the
two-point correlation function. 
These statistical measurements suggest a low value of
the density parameter $\Omega \sim 0.4$ if
optical galaxies are not strongly biased tracers of mass.


\end{abstract}

\section{INTRODUCTION}

Recent developements on extragalactic distance indicators
(Djorgovski and Davis, 1987; Dressler et al., 1987a) allow us to study 
the peculiar galaxy velocity field in the local Universe 
up to $\sim 50-100 h^{-1}$Mpc (see 
Giovanelli, 1997, or Strauss and Willick, 1995 for a review).   
These measurements of peculiar velocities provide direct probes of the 
mass distribution in the Universe and set constraints on 
models of large-scale structure formation. 
By comparing the mass distribution 
implied by the velocity field with the observed distribution of galaxies
the density parameter 
$\Omega$ can be estimated. 
Nevertheless, $\Omega$ may only be determined within the uncertainty of the 
bias parameter $b$ 
($b=1/\sigma_8$ is the inverse of the root mean square mass 
fluctuations in spheres of radius$=8Mpc$ $h^{-1}$)
through the factor $\beta=\Omega^{0.6}/b$.

Bertschinger \& Dekel (1989) 
developed the POTENT method whereby the mass distribution may be 
reconstructed by using the analog of the Bernoulli equation for 
irrotational flows.
This method was  used to analyse the peculiar velocity field
out to 60$h^{-1}$Mpc (Bertschinger et al., 1989). Dekel et al. (1993) 
compared the previously determined velocity field with the 
observed distribution of galaxies concluding that 
$\Omega^{0.6}/b \simeq 1$ provides the best-fitting to the data.
Analysis of the 
velocity tensor
(Gorski, 1988; Groth, Juszkiewics and Ostriker, 1989) 
also provide useful insights
on the velocity field.                        
Zaroubi et al. (1997)
reconstructed the large-scale 
power spectrum from the velocity tensor of the 
MarkIII data and found it consistent with a CDM model with 
$\sigma_{8} \Omega^{0.6} \simeq 0.8$ 
although a different result, $\sigma_{8} \Omega^{0.6} \simeq 0.35$,
is found by Kashlinsky (1997) in a similar analysis.

Relations between root mean square mass fluctuation in a given scale
and density parameter may also be obtained in 
studies of cluster abundances in different cosmological models. 
These analysis place
constraints of the form $\Omega^{\alpha}/b \simeq 0.4-0.6$ with
$\alpha \simeq 0.4-0.6$ in a variety of cold and mixed dark matter models
(see  Eke et al., 1998,  and Gross et al., 1998).
Similarly, studies of redshift space distortions 
of the galaxy two point correlation function also provide a useful 
restriction to the parameter $\beta$,  as for instance
Ratcliffe et al. (1997) who find
$\beta \simeq 0.5$. 
 
In this paper we study the peculiar velocity field  
through a statistical analysis of 
observational data taken from the Mark III catalog.
Data characteristics are presented in section 2.  Section 3 provides an 
outline of the statistical procedure.  In section 4 we analyse the Mark III 
spirals, and in section 5 we test our procedure using mock catalogs
according to different observers in fully
non-linear numerical simulations.  In section 6 we provide a determination
of the parameter $\beta$.

\section {DATA}

We use samples of spiral galaxies taken from
the Mark III catalog 
(Willick et al., 1995; Willick et al., 1996;
Willick et al., 1997)
to analyze the peculiar velocity flow.  
This Catalog lists Tully-Fisher and $D_n-\sigma$ 
distances and radial velocities 
for spiral, irregular, and elliptical galaxies.
For spiral galaxies, the velocity parameter $\eta = Log \Delta V - 2.5$ is 
determined either from HI profiles 
or from optical $H_{\alpha}$ rotation curves.
The Tully-Fisher (TF) relations and their corresponding scatters 
for the different samples of spiral galaxies
are given by Willick et al. (1997) and are shown in Table 1, where the absolute
magnitude $M$ satisfies $M=m-5\log cz$.
The galaxy apparent magnitudes $m$ of the
Tully-Fisher distances are corrected for Galactic
extinction, inclination and redshift
(see Willick et al. 1997 for details).

\begin{table*}
\begin{minipage}{80mm}
\caption{Observations: The Mark III spirals}
\tabskip =1em plus2em minus.5em
\label {Table 1}
\begin{tabular}{cccc}
Subsample  & $N^o$ of Gx. & TF relation & $\sigma_{TF}$  \\
 \noalign {\vskip 10pt}
 \noalign {\hrule}
 \noalign {\vskip 10pt}
 Aaronson et al. Field (1982)& 359 &  $M_H=-5.95+10.29 \eta$ & $0.47$ \\
 Mathewson et al. (1992) & 1355 & $M_I=-5.79+6.8 \eta$ & $0.43$ \\
 Willick, Perseus Pisces (1991) & 383 & $M_r=-4.28+7.12 \eta$ & 0.38 \\
 Willick, Cluster Galaxy (1991) & 156 & $M_r=-4.18+7.73 \eta$ & $0.38$ \\
 Courteau-Faber (1993) & 326 & $M_r=-4.22+7.73 \eta$ & $0.38$\\
 Han-Mould et al., Cl. Gx. (1992) & 433 & $M_I=-5.48+7.87 \eta$ & 0.4 \\
\end{tabular}
\end{minipage}
\end{table*}

The selection bias in the calibration of the forward TF relation
can be corrected once the selection function is known.  
But then the TF inferred distances and the mean peculiar velocities 
are subject to Malmquist bias.
Suitable  procedures to consider these biases, induced both by
inhomogeneities and selection function, 
have been discussed (see for instance
Freudling et al., 1995,  
and references therein) where the spatial distribution,
selection effects and observational uncertainties are realistically modeled
through Monte-Carlo simulations.
We have adopted in our analysis inverse TF distances 
referred to the Cosmic Microwave Background frame
(Willick et al., 1995; Willick et al., 1996;
Willick et al., 1997).
Inverse TF distances overcome distance dependent selection
bias (see for instance Teerikorpi et al., 1998), nevertheless we have also
tested the results using forward TF distances, fully corrected
for Inhomogeneous Malmquist Bias by Willick et al. (1997).

\section{OUTLINE OF THE ANALYSIS}

The different methods applied to infer the distance to a galaxy are subject to
uncertainties due to observational errors as well as 
scatter and systematics of the galaxy parameters.  
Distances are derived from linear relations between absolute magnitudes $M$ 
and physical properties independent of distance 
as for instance the circular velocity $V_c$ in the Tully-Fisher relation,
or the central velocity dispersion in the $D_n-\sigma$ relation. 
Both rms scatter and
posible shifts in the zero-point of the distance relation
should be taken into account
in studies of the peculiar velocity field considering their
strong influence on the results (Padilla, Merch\'an \& Lambas, 1998).

Since peculiar velocities of galaxies are inferred from redshifts and  
independently estimated distances, the effect of a zero-point shift 
would be observed as a systematic motion of a shell of galaxies
proportional to distance.  The effects induced in the velocity field
by the scatter in the distance relations ($\sigma_{DR}$)
depend on the catalog radial gradient which is affected by distance 
uncertainties.

We correct the observed radial gradient by considering a gaussian
distribution of distance uncertainties. 
Therefore the resulting distribution of distance measurements of galaxies
restricted to the same true distance bin $d_a$ is approximately
gaussian centered at $d_a$, with scatter
$\simeq d_a \ \sigma_{DR}$. Here $d_a$ is given in units of $km/s$ and 
$\sigma_{DR}$ corresponds to a distance fraction.  Galaxies 
from other distances $d_b$ ($n_{ab}$) will also contribute
to the measured number of galaxies at $d_a$. 
The corresponding contribution from objects at $d_b$ can be expresed as:
\begin{equation}
n_{ab}=T(d_b-d_a,\sigma)n_b, 
\label{nab}
\end{equation}
\begin{eqnarray}
T(d_b-d_a,\sigma)=\frac{1}{\sqrt{2\pi}\sigma}
\int_{(d_b-d_a)-\Delta/2}^{(d_b-d_a)+\Delta/2} 
e^{\frac{-x^2}{2 \sigma^2}} dx. \nonumber
\end{eqnarray}
where $\sigma=d_b \ \sigma_{DR}$ and
$\Delta=300 km/s$
is the adopted binning of galaxy distances corresponding to shells.
Then, the number of galaxies measured at distance $d_a$ ($n'_a$) takes
into account contributions from all other distances: 
\begin{equation}
n'_a=\sum_{d_b=0}^{d_{max}}n_{ab}
\label{grad}
\end{equation}
where $d_{max}$ is the limiting distance imposed to the catalog.

In order to solve equation \ref{grad},
we define the vectors $N'={n'_a}$ and $N={n_b}$
and the matrix $A={n_{ab}}$. Then 
equation \ref{grad} can be rewritten as
\begin{equation}
N'=A \ N
\end{equation}
$A$ can be inverted to obtain the true number count of
galaxies, $N$, unaffected by the distance estimator scatter taken
into account in the matrix $A$.  
However, a direct inversion
of the matrix (using Gauss method for instance) produces diverging solutions 
for the last components of $N$. This divergence is produced by
accumulation of large errors through the calculation over matrix rows.  
To avoid this problem, 
we used an iterative method in wich we apply $A$ to
$N'$ and obtain $N''$. 
Since $A \sim I$,   
the vector $N_1=N'+(N'-N'')$, will be 
a better approximation to $N$ than $N'$. 
After k iterations, we obtain
\begin{equation}
N_{k+1}=N_k+(N'-N'_k)
\end{equation}  
where we impose the condition $N_k \geq 0$.

This method also accumulates errors in the last components of $N_k$,
but the solutions start to diverge only when more than $3$ iterations
are applied.
The optimal number of iterations was found to be $k=2$ or $3$.
The results of the iterative method have also been checked with 
the mock catalogs analysed in section 5, and were found to be accurate.

\begin{figure*}
\centerline{\psfig{file=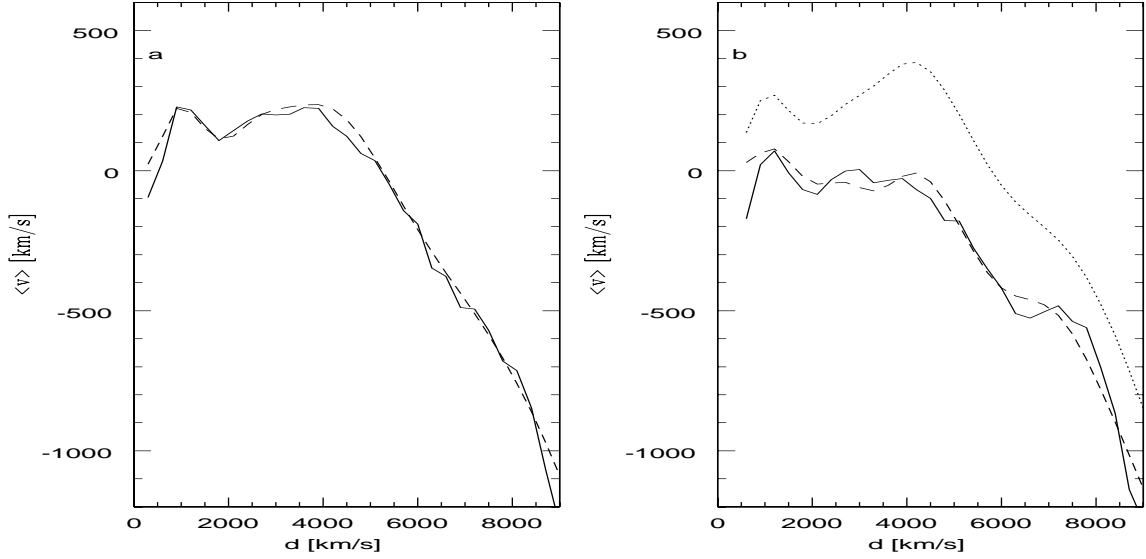,width=16cm,height=8cm}}
\caption{Figure 1a:  Mean radial velocities of shells of galaxies from 
Mark III using inverse TF distances (solid lines) with
$\sigma=0.41$ and $P_0=0.05$. The dashed line corresponds to the model 
results with the same values of $\sigma$ and $P_0$.
Figure 1b:Mean radial velocities 
using forward TF distances corrected for inhomogeneous Malmquist 
Bias (solid lines).  The dashed line corresponds to the model results
with parameter $\sigma_{TF}=0.3$ and $P_0=-0.13$, and  the dotted line to
a model with $\sigma_{TF}=0.4$ and $P_0=0$.
}
\label{fig1}
\end{figure*}

In order to study the effects induced by the TF scatter in the velocity field, 
we calculate the effect of $\sigma_{DR}$ on the mean radial velocity of
galaxies ($v_a$) at distances $d \in [d_a-\Delta,d_a+\Delta]$ 
under the assumption that the shells do not expand nor contract.  
We may write
the apparent mean velocity of the shell $v'_a$ at distance $d_a$ as:
\begin{equation}
v'_a=\frac{1}{n'_a}\sum_{i=1}^{n'_a}v'_i
\end{equation}
where $v'_i$ is the individual velocity of the galaxy $i$.  We recall
the fact that at distance $d_a$ there are contributions from other
distances.  If a galaxy $j$ at real distance $d_b$ is measured 
to be at distance $d_a$, the inferred velocity will be $v'_{j}=v_{j}-(d_b-d_a)$
where $v_{j}$ is the real peculiar velocity of the galaxy $j$.
Finally, if we sort by real distance galaxies  
accidentally in the shell at distance $d_a$, we can rewrite the last
expression as:
\begin{equation}
v'_a=\frac{1}{n'_a}\sum_{d_b=0}^{d_{max}} \sum_{j=1}^{n_{ab}}  (v_{j}-(d_b-d_a))
\end{equation}
If we consider our assumption $v_b=0$ and add the possible presence
of a zero-point shift $P_0$, we find the final expression:
\begin{equation}
v'_a=\frac{1}{n'_a}\sum_{d_b=0}^{d_{max}} n_{ab}  (d_a-d_b+P_0 d_b).
\label{vmed}
\end{equation}

The inputs of this equation are the number count of galaxies as
a function of distance, the scatter 
$\sigma_{DR}$, and the zero-point shift, $P_0$.  
By comparing the measured values $v'_a$ from a catalog with the 
calculated values given by equation \ref{vmed},
we may infer the uncertainties that affected the measured
distances.

A similar deduction can be applied to obtain the apparent root mean
square velocities $\sigma'_{v_a}$ corresponding to 
galaxies in a given shell
at distance $d_a$.  The assumption made here is that
the true root mean square velocity of a shell 
is independent of distance,  
$$\sigma_{v_a}=\sigma_{1dim}$$
This quantity can be calculated from
the following equation:
\begin{equation}
\sigma_{1dim}^2=\frac{\sum_{d_a=0}^{d_{max}} (n'_a \sigma'_{v_a} \ ^2-
		\sum_{d_b=0}^{d_{max}} n_{ab}[d_b-d_a+P_0 d_b]^2)  }
	      {\sum_{d_a=0}^{d_{max}} n'_a}
\label{vquad}
\end{equation}
The inputs of eq. \ref{vquad} are the observed $\sigma'_{v_a}$,
$n'_a$, the distance relation scatter $\sigma_{DR}$, and its zero-point
offset, $P_0$.
The 1-dimensional velocity dispersion,
$\sigma_{1dim}$, is thus directly obtained from radial velocity data. 
The 3-dimensional velocity dispersion $\sigma_v$ is simply
$\sigma_v=\sqrt{3} \sigma_{1dim},$ assuming isotropy.

\section{APPLICATION TO THE MARK III CATALOG} 

The scatter of the MarkIII spiral TF relation (hereafter $\sigma_{TF}$) 
has been extensively
studied (Mo et al., 1997; Willick, 1991; Mathewson, Ford and Buchorn, 1992), 
and similarly uncertainties of the TF zero-point, $P_0$ (Shanks, 1997; 
Willick, 1991). 
The value of $\sigma_{TF}$ for the different spiral galaxy samples 
in the MarkIII catalog is around $0.4$ measured in units of absolute 
magnitude  (Willick, 1991). These authors give a null shift in
the distance estimations with a mean deviation of $\pm 0.07$ in units
of absolute magnitude.  

\begin{figure*}
\centerline{\psfig{file=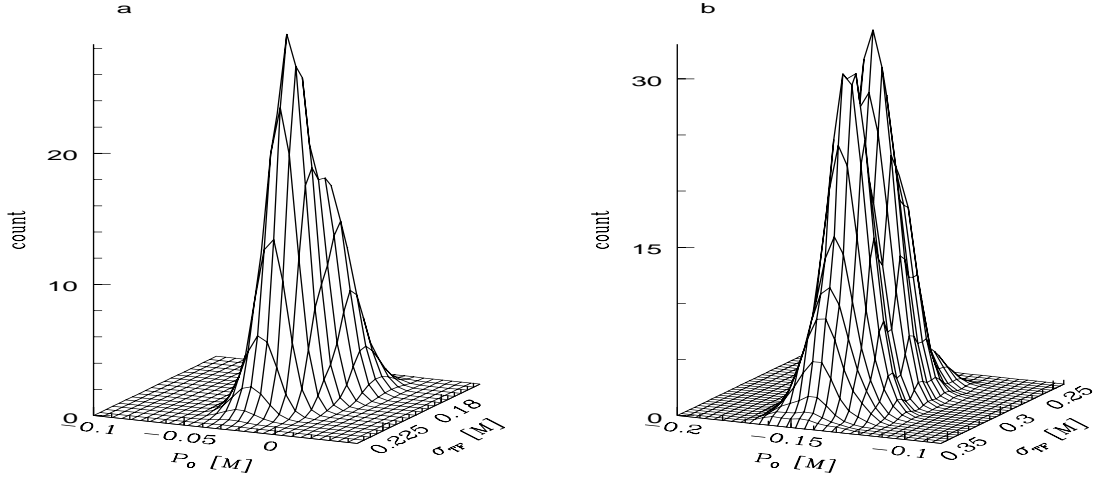,width=16cm,height=8cm}}
\caption{Occurrence of values $P_0$ and $\sigma_{TF}$ by random
resamplings of the sample.   Figure 2a: using inverse TF distances,
Figure 2b: using forward corrected TF distances.
}
\label{fig2}
\end{figure*}     

Since the sample of Mark III spirals has a reasonably smooth 
sky coverage the analysis outlined in the previous section
is suitable for statistical purposes.
We applied eq. \ref{vmed} to the MarkIII spirals
restricted to distances $d < 6000$ $km/s$ since beyond this distance 
the uncertainties make peculiar velocities unreliable.  
We have calculated a 
$\chi^2$ deviation between predicted and observed mean peculiar
velocity of shells $v'_a$ as a
function of $\sigma_{TF}$ and $P_0$.  The best-fitting was obtained for
$$ \sigma_{TF}=0.41, \ \ P_0=-0.05 $$ both in units of absolute magnitude.
Notice that these values of $\sigma_{TF}$ and $P_0$ are consistent
with those quoted in Willick et al. (1997).
Figure \ref{fig1}a and \ref{fig1}b 
show the observed mean velocities and the results of eq. \ref{vmed}
for the values of $\sigma_{TF}$ and $P_0$ obtained using
inverse TF and forward corrected TF respectively.  
In figure 1b it is also shown the predicted $v'_a$ with
no zero-point shift $P_0$ and $\sigma_{TF}=0.4$.
In order to measure the accuracy of the determination of the distance
uncertainties, we apply a $\chi^2$ test to obtain 
$\sigma_{TF}$ and $P_0$ for a large set of catalogs
obtained through bootstrap resampling.  The number of realisations
with resulting $\sigma_{TF}$ and $P_0$, are shown in figures \ref{fig2}a
and \ref{fig2}b for inverse and forward corrected TF distances respectively. 

These results obtained for the inverse TF distances may be compared with
the homogeneous calibration of the Tully-Fisher relation 
corresponding to different
samples of spirals of the Mark III catalog given by 
Willick et al., 1995; Willick et al., 1996; and
Willick et al., 1997; 
which corresponds to 
values of $\sigma_{TF}$ in the range $0.38-0.47$ $mag$ (see Table 1).

\begin{figure}
\centerline{\psfig{file=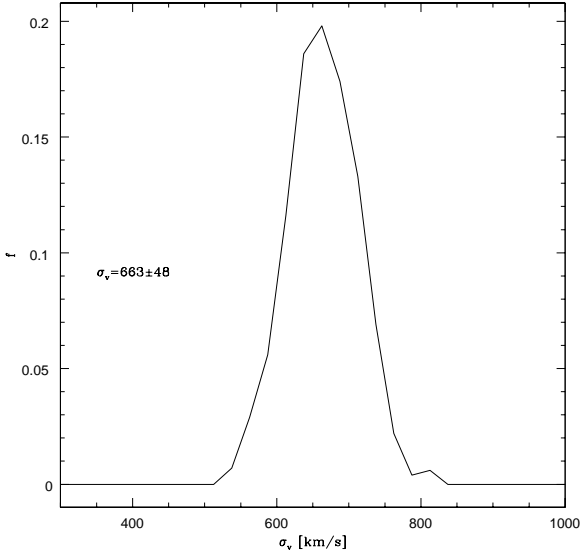,width=8cm}}
\caption{Distribution of $\sigma_v$ for the observational
sample corresponding to bootstrap resampling. }
\label{fig4}
\end{figure}

We have obtained
the galaxy root mean square velocity by application of equation
\ref{vquad} to the sample $d<6000km/s$.
The value obtained is $\sigma_v=(660 \pm 50)$ $km/s$, where
the error was obtained from $1000$ bootstrap resamplings
(Barrow, Bhavsar \& Sonoda, 1984).  The distribution
of $\sigma_v$ derived from the different bootstrap resamplings is shown
in figure \ref{fig4}.

\section{TESTING THE METHOD WITH MOCK CATALOGS }

We test the results of our analysis using  
mock catalogs derived from the numerical simulation.
We adopted a $\Omega=0.5$, $\Omega_{\Lambda}=0$ COBE normalised CDM model
which reasonably reproduces several statistical tests of large-scale
structure such as cluster abundances, correlation functions, etc.
This particular model requires no strong 
bias, so each particle of the simulation corresponds to a galaxy.

We have considered 1000 random observers in the numerical simulation by defining
cones with different positions and orientations in our computational volume.  
We have included the observed strong radial gradient wich 
corresponds approximately to a selection
bias due to a magnitude limit cutoff  in the data to the mock catalogs.
This can be seen from the observed
distribution of absolute magnitudes which is nearly gaussian with
mean $\simeq M^*$ (the knee of the Luminosity Function)
and $\sigma \simeq 1.5mag$.  
Nevertheless, for our statistical purposes
it is not necessary to adopt a Monte-Carlo model using the Galaxy
Luminosity Function in the simulations.  
It suffices to reproduce the observed radial gradient in the
numerical models through a Monte-Carlo rejection algorithm.
Furthermore, we restrict 
the resulting number of particles of the 
mock catalogs to be equal to the number of 
galaxies in the observational sample.

\begin{figure}
\centerline{\psfig{file=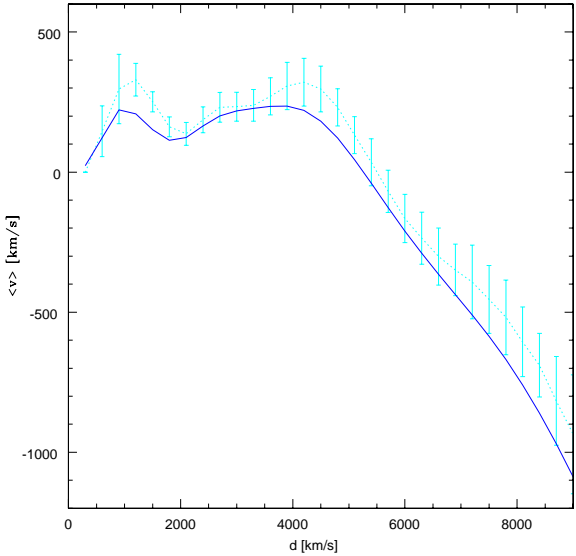,width=8cm}}
\caption{Mean radial velocities of shells according to models (solid line)
and mock catalogs (dashed lines).  Error bars correspond to
fluctuations arising from different observers in the simulations.}
\label{fig5}
\end{figure}     

\begin{figure}
\centerline{\psfig{file=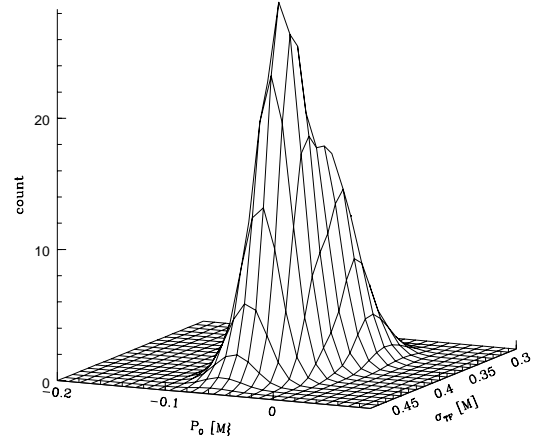,width=8cm}}
\caption{occurrence of values $P_0$ and $\sigma_{TF}$ in random
resamplings of a mock catalog with $\sigma_{TF}=0.41$ and $P_0=-0.05$}
\label{fig6}
\end{figure}     

\begin{figure}
\centerline{\psfig{file=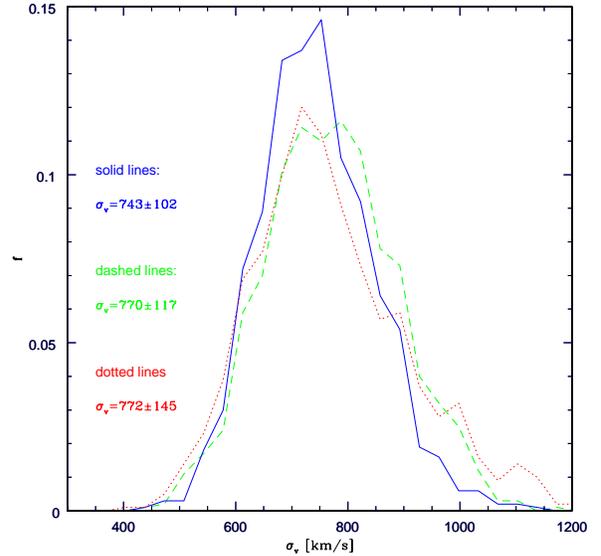,width=8cm}}
\caption{Probability distribution of $\sigma_v$ corresponding
to mock catalogs with 
$\sigma_{TF}=0.3$ and $P_0=0$ (solid lines),
$\sigma_{TF}=0.41$ and $P_0=-0.05$ (dashed lines), and
$\sigma_{TF}=0.4$ and $P_0=-0.2$ (doted lines).
}
\label{fig7}
\end{figure}

Observational errors
in galaxy distance estimates 
were considered assuming gaussian errors in the galaxy absolute magnitudes
of the TF relation with dispersion  $\sigma_{TF}$, corresponding to
relative errors in distances $\frac{\Delta d}{d} \simeq \frac{\sigma_{TF}}{2}$.
Namely, we assign to each particle in the mock catalog a new distance
$d_{new}=d(1+s)$ where $s$ is taken from
a gaussian distribution with dispersion corresponding to the TF uncertainty.  
Then, as the particle peculiar velocities are inferred from the
galaxy redshift and distance, $v_{new}=v-d \ s$.

The N-body numerical simulation  was performed using the Adaptative
Particle-Particle Particle-Mesh (AP3M) code
developed by Couchman (1991). The initial 
condition was generated using the Zeldovich approximation and 
corresponds to the adiabatic CDM power spectrum 
with $\Omega=0.5$ and $\Omega_{\Lambda}=0$. We have adopted the
analytic fit to the CDM power spectrum given by Sugiyama (1995):

\begin{equation}
P(k) \propto \frac{k}{A}
\left( \frac{ln(1+2.34q)}{2.34q} \right) ^{2}
\end{equation}

\noindent where 
$A=[+3.89q+(16.1q)^2+(5.46q)^3+(6.71q)^4]^{1/2}$,
$q=\frac{k}{\Gamma h}$Mpc, $\Gamma=\frac{k\theta^2}
{h \ exp(-\Omega_B- \sqrt{h/0.5} \Omega_B / \Omega)}$, $\theta$ is the
microwave background radiation temperature in units of $2.7K$,
and $\Omega_B=0.0125h^{-2}$ is the value of the baryon density
parameter given by nucleosynthesis.  The normalization of the CDM power
spectrum is imposed by COBE measurements using the value of
$\sigma_{8}$ and $h$ (the Hubble constant in units of $100$ $km/s/Mpc$)
chosen from Table 1 of Gorski et al. (1995) corresponding to 
an age of the universe $t_0 \simeq$ 12 Gyr.
The computational volume is a periodic cube of side length 300 Mpc.
We have followed the evolution of $N=5\times 10^{5}$ particles with a $64^3$
grid and a maximum level of refinements of 4. The resulting mass 
per particle is $2.05 \times 10^{12} h^{-1} M_{\sun} $. 
The initial condition corresponds to redshift $z=10$ and the evolution
was followed using 1000 steps. At the final step ($z=0$) the linear
extrapolated value of $\sigma_{8}$ is compatible with the normalization
imposed by observed temperature fluctuations in the cosmic background.

\begin{figure*}
\centerline{\psfig{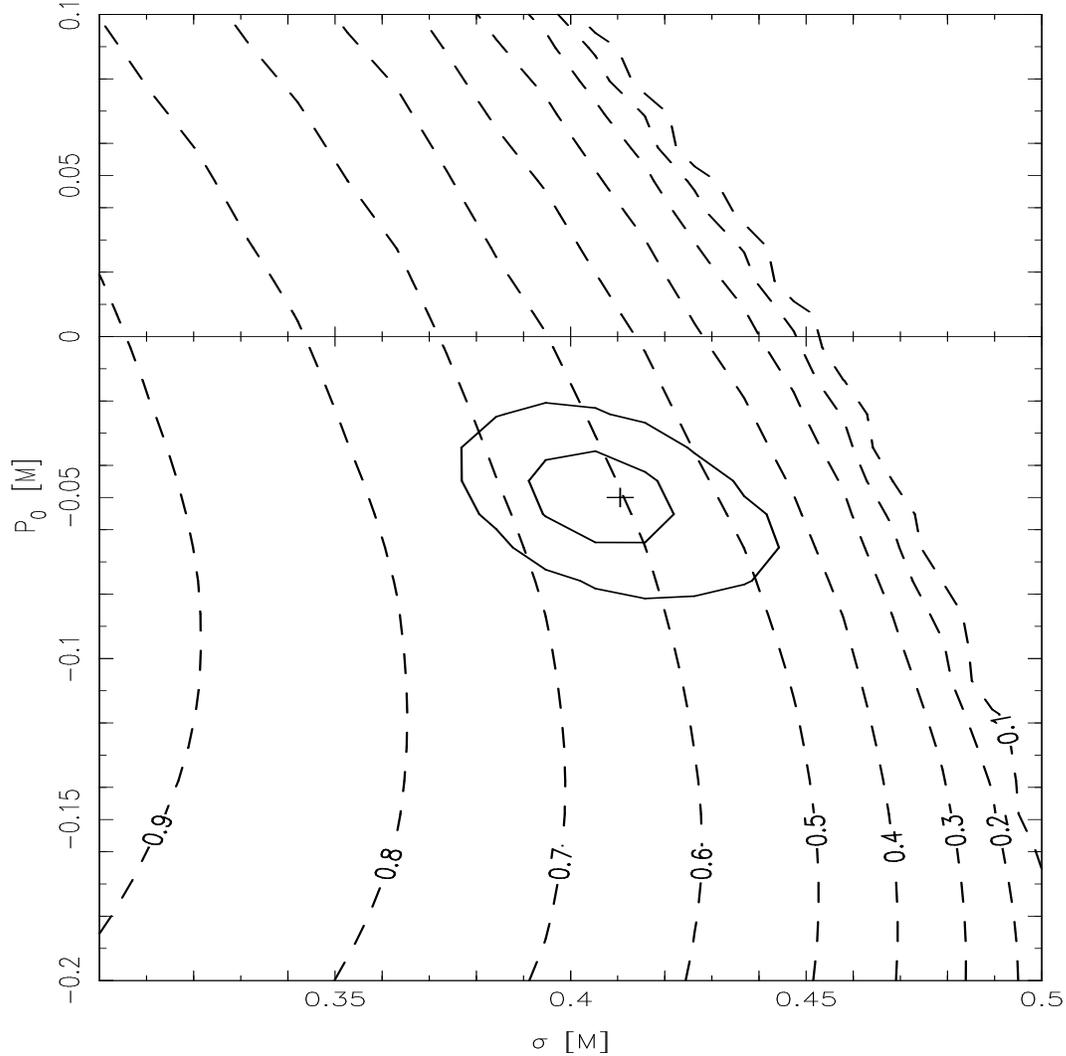}}
\caption{Probability contours corresponding to the observational sample
$d < 6000km/s$. Dashed lines corresponding to different 
values of $\beta$.  The plus sign indicates the best values of
$\sigma$ and $P_0$. }
\label{fig8}
\end{figure*} 

As a test of the ability of equation \ref{vmed} to reproduce
the distorted values of mean velocity, we plot in figure \ref{fig5} 
the resulting $v'_a$ for the average of 1000 mock catalogs, each with
distance uncertaintes included. 
The error bars correspond to the rms deviation of
$v'$ from the different observers in the numerical simulation.
We adopt the values of $\sigma_{TF}$ and $P_0$ used in
the construction of the
mock catalogs to perform the calculation of $v'_a$ 
through equation \ref{vmed}.
It can be seen in the figure the excellent
agreement between the results of the calculation and
the mock catalogs.  
We have applied a $\chi^2$ method to derive $\sigma_{TF}$ and $P_0$.  
We find that our method 
gives accurate estimates of the actual $\sigma_{TF}$ and 
$P_0$ when applied to the mock catalogs. 
In figure \ref{fig6} we show the frequency of $\sigma_{TF}$ and $P_0$ obtained
from bootstrap resampling objects corresponding to a mock catalog with
an imposed scatter $\sigma_{TF}=0.4$ and $P_0=-0.05$ as derived from
the observations.

The values of $\sigma_{TF}$ and $P_0$ inferred from the $\chi^2$ method
can be used to calculate the value of $\sigma_v$ in the mock catalog
through equation \ref{vquad}.  The distributions of inferred 
$\sigma_v$ found  
are shown in figure \ref{fig7} where it can be appreciated the
stability of the method in estimating $\sigma_{1dim}$
under different zero-point offsets $P_0$ and $\sigma_{TF}$.
The actual value of the rms peculiar velocity of the simulation
is $\sigma_v \simeq 780$ $km/s$ which may be compared to the inferred values 
from equation \ref{vquad} for different adopted values of $\sigma_{TF}$
and $P_0$.  For instance, we find
$\sigma_v= (743\pm102)$ $km/s$, $\sigma_v= (770\pm117)$ $km/s$, and  
$\sigma_v= (772\pm145)$ $km/s$ for $\sigma_{TF}=0.3$, $P_0=0$;
$\sigma_{TF}=0.41$, $P_0=-0.05$; and
$\sigma_{TF}=0.4$, $P_0=-0.2$ respectively indicating the ability of our
procedure to derive the rms peculiar velocity irrespectively of
TF scatter and zero-point offset.

We find a zero-point shift $P_0=-0.05$ in our analysis of the 
observations using inverse TF distances.  
It is of interest to test the probability
of occurrence of such a value arising from  
our model assumptions such as $v_a=0$, etc.
We use mock catalogs with imposed $\sigma=0.4$ and 
$P_0 = 0$ in the TF calibration and we test 
the probability of finding different values of zero-point shifts .
We apply a $\chi^2$ method to derive the pair of values
$\sigma_{TF}$ and $P_0$ that provides the best-fitting of equation \ref{vmed}
to the actual values for each mock catalog. We compute the
frequency of occurrence of $\sigma_{TF}$ and $P_0$ for the difference 
observers, and we can estimate the probability of obtaining different
values of zero-point shifts.  
We find that 
a random occurrence of the observed value $P_0=-0.05$ is  
within a standard deviation, consistent with Willick et al. (1997) estimate.

\section{DETERMINATION OF THE $\beta$ PARAMETER}

The 3-dimensional velocity dispersion $\sigma_v$ is directly
related to the quantity $\beta=\Omega^{0.6}/b$ through the relation 
\cite{peebles}
\begin{equation}
\sigma_v = \left(\frac{H a f}{b}\right)^2 \int_0^{\infty}y\xi (y)dy
\label{ome}
\end{equation}
where $H$ is the Hubble constant, $a$ is the expansion factor of the universe,
$y$ is expressed in $Mpc$, $\xi(y)$ the galaxy spatial correlation function,
$f\simeq \Omega^{0.6}$ the rate of growth of inhomogeneities, and $b$ is the
linear bias factor.
We express $y$ in units of 
$Mpc$ $h^{-1}$, therefore $H=100 km/s/Mpc$ and $a=1$.

We estimate $\Omega^{0.6}/b$  
from the inferred value of 3-dimensional root mean square peculiar velocity.
We adopt the power-law fit to the galaxy spatial correlation function 
estimated by Ratcliffe et al. (1997):
$$
\xi(r)= \left\{
\begin{array}{ll}
 (\frac{r}{r_0})^{-\gamma} & \mbox{if $r \leq 50$ $Mpc$} \\
 0			   & \mbox{if $r > 50$ $Mpc$} 
\end{array}\right.
$$
with $r_0=5.1$ $Mpc$ and $\gamma=1.6$.

We calculate $\beta$ from equation \ref{ome}  
using equation \ref{vquad} to express $\sigma_v$ and therefore $\beta$ in
terms of the parameters $\sigma_{TF}$ and $P_0$. 
In figure \ref{fig8}
we show equal $\beta$ contours in the $\sigma-P_0$ plane
for our sample of Mark III spirals with distances
$d < 6000$ $km/s$.  
Also shown in this figure are the 1 $\sigma$ and 2 $\sigma$ contour levels
corresponding to the frequency of inferred $\sigma_{TF}$ and $P_0$
from bootstrap resamplings of the observational data set.  
The corresponding result is $\beta = 0.60^{+0.08}_{-0.05}$.

The true uncertainty of the global value
$\beta = \Omega^{0.6}/b$ derived from a catalog of peculiar velocities
will be grater than 
that obtained from bootstrap resampling of the data due to cosmic
variance.
A suitable value of the uncertainty in $\beta$ can be estimated
from the rms values of this parameter 
derived from the mock catalogs.  According to our analysis
$\Delta \beta \simeq 0.1$
for a limiting distance $d_{max}<6000$.
Thus, adding in quadrature both errors, we find $\beta = 0.60^{+0.13}_{-0.11}$  
for $d<6000km/s$.

Thus, the observed peculiar velocity field
is inconsistent with a critical density universe if 
optical galaxies trace the mass.

\section{CONCLUSIONS}

We have developed a method for the 
analysis of the peculiar velocity field inferred from peculiar
velocity data and we apply this procedure to the spirals of the
Mark III catalog.  
We estimate 
optimal values of inverse Tully-Fisher scatter and zero-point offset for
a sample of the catalog with limiting distance $d_lim = 6000 km/s$.
We derive the 3-dimensional rms peculiar velocity 
of the galaxies $\sigma_v = (660\pm50) km/s$ where the uncertainty
has been obtained through bootstrap resampling of the data.  

In our model, the shells have not a net mean
radial motion, and the mean square velocities of galaxies in
different shells are described by a unique number $\sigma_v$.
The comparison with mock catalogs derived from numerical simulations
shows that these are reasonable hypotheses that allow to obtain
physical characterizations of the nearby universe from observations.
We have shown that corrected TF distances require a
large $P_0 \simeq -0.15mag$ so that  caution should be taken
when they are used in statistical analysis.  

We use mock catalogs derived from 
numerical simulations of CDM models considering 
measurement uncertainties and sampling variations  
to check our statistical analysis.
We find a general good agreement between the results of the calculations
and those measured in the mock catalogs.  
The spread of $\sigma_v$ measurements from different 
observers in the numerical
simulations may be added in quadrature to the bootstrap
resampling errors to provide a more reliable estimate of the
uncertainty in $\sigma_v$.
We infer $\sigma_v=660\pm70$ $km/s$, and  we conclude that  
$\beta \simeq \Omega^{0.6}/b = 0.60^{+0.13}_{-0.11}$.

Estimates of the parameter $\beta$ from other analysis  such as 
studies of redshift space distortions 
of the galaxy two point correlation function provide   
similar values  $\beta \simeq 0.5$ (Ratcliffe et al. 1997).
Moreover, the confrontation of observed cluster abundances with prediction of
different cosmological models put 
constraints of the form $\Omega^{\alpha}/b \simeq 0.4-0.6$, values 
consistent with our determinations.

\section{ACKNOWLEDGEMENTS}

We have benefitted from helpful discussions with Jim Peebles and David 
Valls-Gabaud.
This work was supported by the Consejo de Investigaciones Cient\'\i ficas y
T\'ecnicas de la Rep\'ublica Argentina, CONICET, the Consejo de
Investigaciones Cient\'\i ficas y Tecnol\'ogicas de la Provincia de C\'ordoba,
CONICOR, and Fundaci\'on Antorchas, Argentina.

{}

\end{document}